\documentclass[prb,twocolumn,notitlepage]{revtex4-1}
\usepackage[hidelinks]{hyperref} 
\usepackage{amsmath}
\usepackage{lmodern}
\usepackage{amssymb}
\usepackage{epsfig}

\def \d {\mathrm{d}}
\def\pd{{\phantom\dagger}}
\newcommand{\be}{\begin{equation}}
\newcommand{\ee}{\end{equation}}
\newcommand{\beA}{\begin{equation}\begin{aligned}}
\newcommand{\eeA}{\end{aligned}\end{equation}}

\begin{document}
\title{The fate of quantum shock waves at late times}
\date{\today}
\author{Thomas Veness}
\author{Leonid I. Glazman}
\affiliation{Department of Physics, Yale University, New Haven, Connecticut
06520, USA}
\begin{abstract}
Shock waves are an ubiquitous feature of hydrodynamic theories.
Given that fermionic quantum many-body systems admit hydrodynamical
descriptions on length scales large compared to the Fermi wavelength, it
is natural to ask what the status of shock waves is in such systems.
Free fermions provide a solvable yet non-trivial example, and
here we generalise to include generic (non-integrable) weak interactions to
understand how a shock wave decays and changes its shape well after forming.
\end{abstract}

\maketitle

\section{Introduction} 
The notion of shock waves is well established in classical
hydrodynamics\cite{witham}.
If the fluid velocity is an increasing function of density, then any smooth
density profile with a local maximum will eventually form a shock wave: a
physical quantity becomes non-analytic as a function of spatial coordinates.
In fermionic systems, there is a tension between this singular
behaviour and the dispersive broadening one may expect at the level of
single-particle quantum mechanics.

The nature of shock waves in the context of free fermions has been the subject
of previous theoretical investigation \cite{Protopopov, BettelheimGlazman,
AbanovWiegmann1,AbanovWiegmann2,Bettelheim}.
In a particular classical limit, the formation of a shock wave is exhibited as
a non-analyticity in the density $\rho(x)$.
Semi-classical corrections modify this by smoothing out the behaviour at the
shock front through the introduction of quantum ripples.
It should be emphasised that prior work has focussed on times close to the
formation of the shock, where the ripples may be significant across the
entirety of the structure associated with the shock.
In Section~\ref{sec:freefermions}, we recapitulate these results and observe
that for times well after shock formation a parametrically large spatial region
has only negligible quantum corrections to the density.

Recently, the topic of generalised
hydrodynamics\cite{BertiniPRL,DoyonDubail,DoyonYoshimura,CastroAlvaredo} has
led to significant progress in understanding the dynamics of quantities such as
the density for integrable systems.
The consequences of generic (i.e. non-integrable) interactions are not
clear from this picture, however.
In this paper we linearise a Boltzmann equation and use single particle decay rates to
describe the effect of interactions.
This technique allows us to investigate the shock wave at all spatial scales,
excluding only a small spatial region affected by quantum corrections, and is
valid for times well beyond that of shock formation.

The kinetic theory developed in Section~\ref{sec:interactions} allows us to
find the deformation of the spatial distribution of the density caused by
relaxation. Despite the exponential decay of the number of fermions forming the
shock wave, a substantial section of the shock wave retains its profile.

We present the final conclusions in Section~\ref{sec:conc}, where we associate
the dissolution of the shock wave with the interplay between the
quantum-mechanical dispersion and the quasiparticle kinetics.
\section{Shock waves for free fermions}
\label{sec:freefermions}
In order to present a self-contained discussion, we begin by recapitulating and
extending some results from Ref.~\onlinecite{BettelheimGlazman}.
The question we wish to address is the following: for a system of spinless
fermions, given an initial density profile
\be
\rho(x) 
=
\frac{k^0}{\pi} + \frac{k^m-k^0}{\pi}f\left( \frac{x}{w} \right),
\label{eq:rhoinitial}
\ee
how does the density evolve as a function of time? In Eq.~(\ref{eq:rhoinitial})
$k^0/\pi$ is a background density corresponding to a uniform Fermi sea, and
$\left( k^m-k^0 \right)/\pi$ is the height of an isolated, smooth perturbation
with profile $f(s)$. 
This has a single maximum at $f(0)\sim 1$ and $\lim_{|s|\to\infty}f(s)=0$,
varying on the scale $s\sim 1$ (the perturbation is of extent $\sim w$).
We restrict to the scenario where the height of the perturbation is small $k^m
- k^0 \ll k^0$, and the number of particles contained in the perturbation is
large $\Delta N \sim \left( k^m-k^0 \right)w \gg 1$. 
$\Delta N$ will be our large parameter for a semi-classical treatment.

The small height of the perturbation implies that excitations are confined to
be particles and holes in the vicinity of the Fermi points. 
This gives us a well-defined notion of right- and left-movers.
We initially consider the case of free fermions with a parabolic dispersion
relation and mass $m$ as given by the Hamiltonian
\be
H=\sum_p \frac{p^2}{2m} \psi^\dagger_p \psi^\pd_p,
\label{eq:Hquad}
\ee
where $\psi^\dagger_p$ and $\psi^\pd_p$ are fermionic creation/annihilation
operators at momentum $p$ and obey the standard anti-commutation relations, and
we set $\hbar=1$ throughout.

We are interested in a semi-classical description of the problem, and so
introduce the Wigner function, defined by
\begin{multline}
W(x,k,t)\equiv \int \d y e^{-iky} \\
 \langle \Psi| e^{iHt}  \psi^\dagger\left( x-\frac{y}{2} \right)
 \psi\left( x+\frac{y}{2} \right) e^{-iHt} |\Psi\rangle,
\label{eq:WignerDef}
\end{multline}
where $|\Psi\rangle$ is the initial state at $t=0$.
The Wigner function is useful for a number of reasons: it allows us to perform
a controlled semi-classical approximation with large parameter $\Delta N$,
it provides simple access to the density, given by
\be
\rho(x,t) = \int \frac{\d k}{2\pi} W(x,k,t),
\label{eq:rhodef}
\ee
and finally, for $H$ given by Eq.~(\ref{eq:Hquad}), $W(x,k,t)$ obeys the simple
linear differential equation
\be
\left( \partial_t + \frac{k}{m} \partial_x \right) W(x,k,t) = 0.
\ee
This results in the time-evolved Wigner function having the form $W(x,k,t) =
W\left( x-kt/m,k,0 \right)$.  
\subsection{Classical picture}
It is a natural ansatz that, due to the smooth variation of the density on the
scale of the Fermi wavelength, the Wigner function of Eq.~(\ref{eq:WignerDef})
may be described by a ``local Fermi surface'' i.e.
\be
W_0(x,k,0) \equiv \theta(k_F(x)-k) \theta(k_F(x)+k).
\ee
Right- and left-movers separate on a timescale $t_{LR} \sim \frac{mw}{k^0}$.
We therefore choose to ignore left-movers with no loss of generality, and
simplify the above to
\be
W_0(x,k,0) \approx \theta(k_F(x)-k) \theta(k^0+k) \label{eq:stepfunctionWigner}
.
\ee
This leads to the implicit equation 
\be
k_F(x,t)=k_F( x-k_F(x,t)t/m,0),
\ee
which gives rise to multi-valued solutions on a time-scale $t_S \sim
\frac{mw}{k^m-k^0}$. This is consistent with ignoring left-movers, as
$t_{LR} \ll t_S$.
The region where $k_F(x,t)$ is multi-valued exists between the front of the
shock, which we denote $x_+(t)$; and the back of the shock $x_-(t)$.
Formally $x_\pm(t)$ are the two solutions of $\partial_k x_F(k,t)=0$ where
$x_F(k,t)$ satisfies $k_F\left( x_F(k,t),t \right)=k$.
At $t=t_S$ these two solutions coincide.
For $t-t_S\ll t_S$, $x_+(t) - x_-(t) \sim w \left( \frac{t-t_S}{t_S} \right)^2$.
For $t\gg t_S$ the difference between them (i.e. the extent of the shock) grows
linearly in time as $x_+(t) - x_-(t) \sim (k^m - k^0) t$. 

\begin{figure}
		\includegraphics[width=0.95\linewidth]{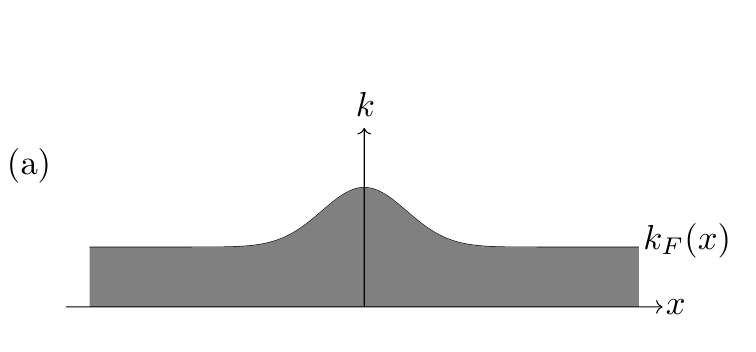}\\ \ \\
		\includegraphics[width=0.95\linewidth]{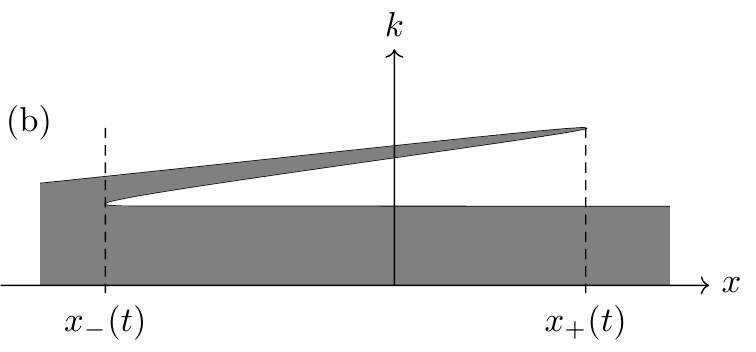}\\ \ \\
		\includegraphics[width=0.95\linewidth]{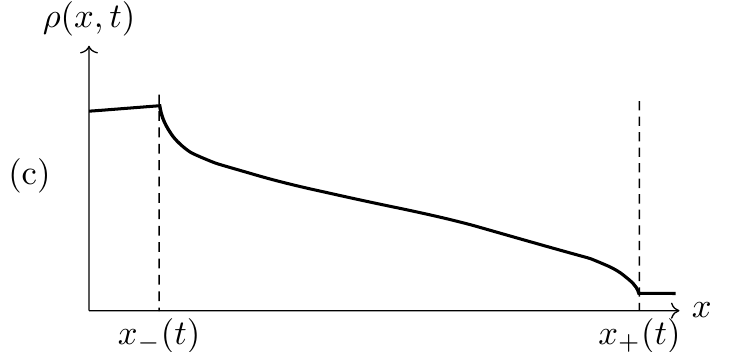}
\caption{
	Schematic for free fermions in the classical limit $\Delta N \to \infty$,
	showing
	(a) the Wigner function for $k>0$ and $t=0$,
	where in the shaded region the Wigner function is 1, and vanishes
	elsewhere; 
	(b) the Wigner function for $k>0$ and $t>t_S$: the shock wave
	has now formed between $x_-(t)$ and $x_+(t)$;
	(c) the density $\rho(x,t)$ corresponding to $t>t_S$, given by
	integrating (b) in the $k$ direction as in Eq.~(\ref{eq:rhodef}). Due to
	taking the classical limit, non-analytic behaviour is observed at
	$x_-(t)$ and $x_+(t)$.
\label{fig:initialWigner}}
\end{figure}

Between the points $x_-(t)$ and $x_+(t)$, $k_F(x,t)$ has three branches which we will denote
$k_F^{(0)}(x,t) \leq k_-(x,t) \leq k_+(x,t)$.
It is evident from Fig.~\ref{fig:initialWigner} that near $x_+(t)$ the density
acquires square-root behaviour in $x_+(t)-x$, and so within the ansatz of
Eq.~(\ref{eq:stepfunctionWigner}) a non-analyticity in the density arises. 

\subsection{Semi-classical corrections}
The main result of Ref.~\onlinecite{BettelheimGlazman} is to quantify how, for
a specific form of initial state $|\Psi\rangle$, including the leading
semi-classical correction rounds off the non-analytic behaviour. 
We begin from the same point, specifying the initial state as
\be
|\Psi\rangle = e^{i \int \d x \rho^R(x) \Phi(x)} |0\rangle
.
\label{eq:psi}
\ee
Here $\rho^R(x)$ is the density associated with right-movers, $|0\rangle$ is
the (translationally invariant) ground state with Fermi momentum $k^0$, and
$\Phi(x)$ is a smooth function corresponding to a density $\rho(x) = k^0 /\pi +
\Phi'(x)/(2\pi)$ i.e.  $\Phi'(x) \leftrightarrow k_F(x)-k^0$.
This state has convenient analytic structure, and is experimentally relevant in
terms of being preparable by a sudden large
perturbation\cite{CobdenMuzykantskii,AbanovWiegmann2}.

Considering only right-movers, standard bosonisation techniques give an
explicit integral representation for the Wigner function at $t=0$ of
\be
W(x,k,0) = \int \d y\, e^{i(k^0-k)y} \frac{e^{i(\Phi(x+y/2) - \Phi(x-y/2))}}{2 \pi i (y+i0^+)}.
\label{eq:Wt0}
\ee
Performing a gradient expansion of $\Phi$ in the exponent, it is clear that
retaining only the linear-in-$y$ term leads to the step-function approximation
of Eq.~(\ref{eq:stepfunctionWigner}). 
This approximation is justified in the $\Delta N\to \infty$ limit, where the
ansatz of Eq.~(\ref{eq:stepfunctionWigner}) as describing
Eq.~(\ref{eq:WignerDef}) is exact.
Keeping the $y^3$ term in Eq.~(\ref{eq:Wt0}) amounts to including
semi-classical corrections.
The excess density may be expressed in terms of the distance from the front of
the shock as
\beA
\frac{\delta \rho(x,t)}{k^m - k^0} 
&
\equiv
\frac{
 \langle \Psi|
\rho(x,t)|\Psi\rangle - \langle 0 | \rho(x,t)|0\rangle
}{k^m-k^0} \\
 &= 
\frac{t_s}{t \Delta N ^{1/3}}
\left[ 
	{\rm Ai}' \left( \frac{  x_+(t)-x}{\ell(t)}\right)^2 
\right. \\
&\ \ \ \left.
	- \frac{ x_+(t)-x }{\ell(t)} {\rm Ai}^2\left( \frac{x_+(t)-x}{\ell(t)} \right) 
 \right]
 ,
\label{eq:dxrhoairy}
\eeA
where the length-scale $\ell(t)$ is given by
\beA
\ell(t)  &= \left( \frac{\partial_k^2 x_F(k,t) |_{k_F(x_+(t),t)}}{2} \right)^{1/3} \\
& \sim \frac{1}{2^{1/3} \left( \Delta N \right)^{2/3}}
\frac{t}{t_S}w,
\label{eq:elldef}
\eeA
and ${\rm Ai}$ is an Airy function\cite{BettelheimGlazman}.
Note that the previous work focussed on times shortly after the formation of
the classical shock ($t-t_S \ll t_S$, $x_+ - x_- \ll w$), and identified the
small parameter $1/\Delta N^{2/3}$ required for the classical description to be
valid.
The main result of Ref.~\onlinecite{BettelheimGlazman} is that the scaling
function of Eq.~(\ref{eq:dxrhoairy}) gives a good description of the ripples
for the entire interval $x_+-x_-\lesssim w$ until times $t - t_S \sim t_S$.
In fact, as shown in Appendix~\ref{app:bulk}, Eq.~(\ref{eq:dxrhoairy})
continues to give a good description at all times $t\gtrsim t_S$ for $x_+(t) -
x \ll x_+(t) - x_-(t)$.

Because the extent of the shock grows linearly with time for $t\gg t_S$, it
will be useful to introduce a dimensionless parameter measuring the distance
from the front of the shock:
\be
\lambda(x,t) \equiv \frac{x_+(t) - x}{x_+(t) - x_-(t)}.
\ee
In terms of this dimensionless variable, we may express the asymptote of
Eq.~(\ref{eq:dxrhoairy}) for $\lambda>0$ as
\be
\frac{ \delta \rho[\lambda,t]}{\left( k^m - k^0 \right)}
\approx
\frac{\sqrt{\lambda}}{\pi}
\frac{t_S}{t}
\left( 
1+
\frac{ \sin \left( \frac{2}{3} (\lambda/\lambda_{\rm cr})^{3/2} \right)}{2\lambda/\lambda_{\rm cr}}
 \right)
,
\label{eq:subleading}
\ee
where the crossover scale $\lambda_{\rm cr} = \frac{1}{(\Delta N)^{2/3}}$.
For the regime $1 \gg \lambda \gg \lambda_{\rm cr}$,
semi-classical corrections are negligible and the simple step-function of
Eq.~(\ref{eq:stepfunctionWigner}) captures the essential physics. 
In other words, although the spatial window where quantum corrections are
appreciable grows with time, it is parametrically small on the length-scale of
the shock.

How this window changes over time can be understood simply: the length-scale
for quantum corrections $\ell(t)$ grows linearly in time.  At times $t-t_S \ll
t_S$, the extent of the shock is small as $x_+(t) - x_-(t) \sim w\left(
t-t_S\right)^2 / {t_S}^2$ and the quantum corrections are significant.
However, at late times $t-t_S\gg t_S$, while the length-scale $\ell(t)$ grows
linearly in time, so too does the extent of the shock and we have that
$\frac{x_+(t)-x_-(t)}{\ell(t)} \sim \Delta N^{2/3}$ i.e. the ripples are
squeezed into a fraction $\Delta N^{-2/3}$ of the shock. 
Therefore quantum corrections are most significant around the time $t=t_S$,
when the shock nucleates. 
For $t-t_S\gg t_S$, the fraction of the shock smeared by quantum
fluctuations $\lambda_{\rm cr}$ remains finite and independent of time.

We wish to add small, generically integrability-breaking interactions to this
picture. 
Having established the regime within which semi-classical corrections are
small, restricting to this will allow us to make further analytic progress.

\section{Adding generic interactions}
\label{sec:interactions}
We wish to understand how adding interactions changes the behaviour at late
times, restricting to small interactions such that the shock structure of free
fermions can become established before decay processes start taking effect.
Well after the formation of the overhanging profile of
Fig.~\ref{fig:initialWigner}(b), particles above the Fermi surface will begin
to relax towards lower energies. 
This modifies the Wigner function from the step-like behaviour of
Fig.~\ref{fig:initialWigner}, and it will generically be non-zero for $k^0< k <
k_-(x,t)$ and $x_-(t) < x < x_+(t)$. 
We will examine how to modify the free fermionic description to account for
integrability-breaking interactions, and how this changes the evolution of the
density as a function of time.

In one dimension, two-particle collisions do not redistribute energy and
momentum.
Generic interactions permit 3-particle collisions, which leads to the
relaxation of excited states\cite{Matveev2}.
We denote the decay rate for single-particle excitations over the Fermi sea
with momentum $k>k^0$ by $\Gamma(k)$.
We wish to incorporate this decay rate, with characteristic magnitude
$\Gamma(k^m)$, into our description of the time-evolution of the shock.
By working explicitly in the regime where $\Gamma(k^m) t_S \ll 1$ the shock
profile is established before decay processes become important.
We also require $t\gg t_S$, $1\gg \lambda \gg \lambda_{\rm cr}$ such that we may
dispense with ripples.
Accordingly, one may view the Wigner function as the distribution function in
the classical limit, $f(x,k,t)$. In the absence of integrability,
three-particle collisions lead to a redistribution of the occupied states, and
this is captured by the kinetic (Boltzmann)
equation\cite{landafshitzsp,Matveev1}
\be
\left( \partial_t + \frac{k}{m} \partial_x \right)f(x,k,t)
=
I\left\{  f  \right\}
,
\ee
where $I$ is a three-particle collision integral.

We wish to evaluate how the shock structure fades on times $t \gtrsim
1/\Gamma(k^m)$. 
At these long times, the variation of the spatial structure is smooth on the
scale of the Fermi wavelength.
Intuitively, interactions will lead to a decay of $f(x,k,t)$ at ``high
energies'' (i.e. for $k$ between $k_-(x,t)$ and $k_+(x,t)$), which will act as
a source for ``low energies'' ($k$ between $k^0$ and $k_-(x,t)$).  Linearising
the collision integral $I$ gives an equation of the form
\be
\left( \partial_t + \frac{k}{m} \partial_x \right)f(x,k,t)
=
J(x,k,t)
-\Gamma(k) f(x,k,t).
\label{eq:boltzmann}
\ee
Here $J(x,k,t)$ represents ``high energy'' particles decaying and acting as a
source for $f(x,k,t)$, and $\Gamma(k)$ is the aggregate of decay processes from
momentum $k$ to lower energies.
In terms of the decay rate from $p$ to the interval $[k,k+\delta k]$, which we
denote $\mathcal{W}_{p\to k}\delta k$, $J(x,k,t)$ and $\Gamma(k)$ are given by
\beA
J(x,k,t) &= \int_k^{k_+(x,t)} \d p f(x,p,t) \mathcal{W}_{p\to k} , \label{eq:Jform}\\
\Gamma(k) &= \int_{k^0}^k \d p\, \mathcal{W}_{k\to p}.
\eeA
Formally, Eq.~(\ref{eq:boltzmann}) is a linear integro-differential equation.
We decompose $f(x,k,t)$ into ``low-energy'' (below $k_-(x,t)$) and
``high-energy'' (between $k_-(x,t)$ and $k_+(x,t)$) pieces:
\be
f(x,k,t) =
\begin{cases}
f_{\rm low}(x,k,t), & k^0<k\leq k_- \\
f_{\rm high}(x,k,t), & k_- < k \leq k_+
\end{cases}
.
\ee
If we focus on the region in $k$ between $k_-(x,t)$ and $k_+(x,t)$, there is no
source for particles: $J(x,k,t)=0$ for $k_-(x,t) < k < k_+(x,t)$.
In the accepted approximation $f_{\rm high}(x,k,t)$ satisfies the equation
\be
\left( \partial_t + \frac{k}{m}\partial_x \right)f_{\rm high}(x,k,t) = -\Gamma(k) f_{\rm high}(x,k,t).
\label{eq:dfhigh}
\ee
The initial conditions
are defined by the free evolution within the time frame
 $t_S \ll t \ll \frac{1}{\Gamma(k^m)}$.
 The corresponding solution of Eq.~(\ref{eq:dfhigh}) is
\begin{multline}
f_{\rm high}(x,k,t) =\\ \theta\left( k-k_-(x,t) \right) \theta\left( k_+(x,t)-k \right) e^{-t \Gamma(k)}.
\end{multline}
This is simply the result of Eq.~(\ref{eq:stepfunctionWigner}) augmented with
the finite lifetime of fermions above the Fermi surface.
Immediately below this region, $k\lesssim k_-(x,t)$, the only contribution to
the source term $J(x,k,t)$ in Eq.~(\ref{eq:boltzmann}) comes from $f_{\rm
high}(x,k,t)$. 
It is therefore appropriate that
Eq.~(\ref{eq:Jform}) may be approximated by
\beA
J(x,k,t)  &\approx
\int_{0}^{k_+(x,t)} \d p\,
f_{\rm high}(x,p,t) \mathcal{W}_{p\to k}
\\
&= \int_{k_-(x,t)}^{k_+(x,t)} \d p\,
 e^{-t \Gamma(p)} \mathcal{W}_{p\to k}
.
\label{eq:Jfhigh}
\eeA
We will comment upon the consistency of this approximation at the end of this
section.
This approach means that $J(x,k,t)$ and $\Gamma(k)$ are independent of
$f(x,k,t)$, and the solution of Eq.~(\ref{eq:boltzmann}) for $k^0<k<k_-(x,t)$
is easily verified as 
\begin{multline}
f_{\rm low}(x,k,t)
=  \\
 \int_0^t \d t' 
e^{-(t-t') \Gamma(k)} 
J\left( x-k(t-t'),k,t' \right) 
.
\label{eq:ffull}
\end{multline}
This corresponds to integrating over all contributions from modes which are
sourced by the term $J$, and also allows for decay. 

Concretely, we now consider a Hamiltonian of the form
\begin{multline}
H = \sum_p \frac{p^2}{2m} \psi^\dagger_p \psi_p   \\
+
\frac{1}{2L}
\sum_{ \substack{ q\neq 0 \\
		\alpha = R,L }
	}
\left( 
V_q \rho^\alpha_q \rho^\alpha_{-q}
+
2U_q \rho^R_q \rho^L_q
 \right)
,
\end{multline}
where $\rho^{R/L}_q = \sum_k {\psi^{R/L}_{k-q}}^\dagger {\psi^{R/L}_{k}}$ is
the Fourier component of the density operator for right- and left-movers
respectively.
It is therefore sensible to consider $\mathcal{W}_{p\to k}$ (and accordingly
$\Gamma(k)$) as that given by the single-particle decay rate of
Ref.~\onlinecite{PKKG}.
In terms of the decay rate $\Gamma(k^m)$ at $k^m$ we can express
\beA
{\mathcal{W}_{p\to k}} &= c \frac{\left( k-k^0 \right)^2 \left( p-k \right)^5}{\left( k^m-k^0 \right)^8} \Gamma(k^m),\\
\frac{\Gamma(p)}{\Gamma(k^m)} &= \frac{\left( p-k^0 \right)^8}{\left( k^m - k^0 \right)^8},
\label{eq:decayrate}
\eeA
with corrections suppressed by factors of $(p-k^0)/k^0 \ll 1$.
Here $c=168$ is a normalisation constant such that $\int_{k^0}^{k^m} \d k
\mathcal{W}_{k^m\to k} = \Gamma(k^m)$.
In evaluating Eq.~(\ref{eq:Jfhigh}), we assume that $t\gg t_S \sqrt{\lambda}$,
and may replace the exponential and the rate $\mathcal{W}_{p\to k}$ by their
averages in the interval:
\be
J(x,k,t)
\approx
\left(
k_+(x,t) - k_-(x,t)
\right)
e^{-t \Gamma(\bar{k}) }
\mathcal{W}_{ \bar{k} \to k}
,
\label{eq:avg}
\ee
where $\bar{k} = \frac{1}{2} \left(  k_+(x,t) + k_-(x,t) \right)$.
and vanishes otherwise. To ease notation, we set $m=1$ for the remainder of the
paper.

\begin{figure}
\includegraphics[width=0.95\linewidth]{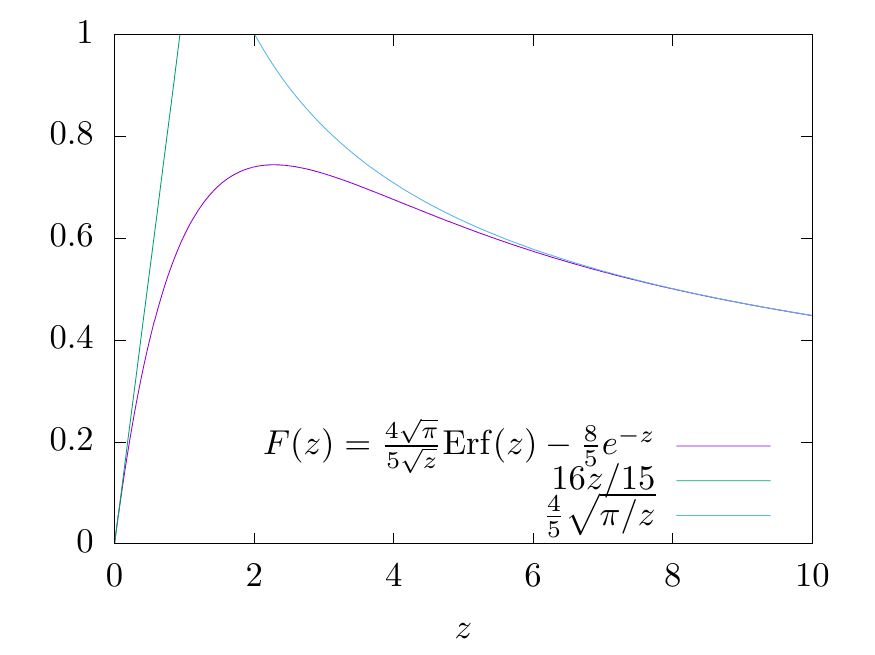}
\caption{Scaling function $F(z)$ of Eq.~(\ref{eq:Fcorrection}) describing correction to $\delta \rho_{\rm high }$. The asymptotes of $F(z)$ are shown for  $z\ll1$, where the behaviour is linear, and $z\gg 1$, where $F(z)$ falls off as $z^{-1/2}$.  }
\label{fig:F}
\end{figure}

\begin{figure}
\includegraphics[width=0.95\linewidth]{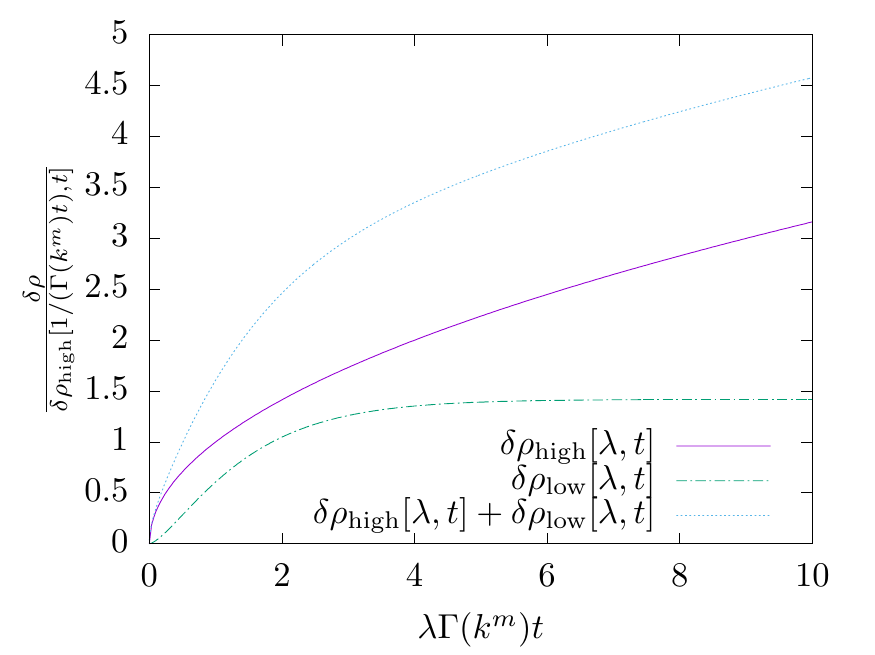}
\caption{
	Plot of the contributions to the shock wave density $\delta \rho=\delta
	\rho_{\rm low} + \delta \rho_{\rm high}$ appearing in
	Eq.~(\ref{eq:combinedResult}).  
	The contribution coming from lower energies, 
	$\delta \rho_{\rm low}$, is comparable to $\delta \rho_{\rm high}$
	when $\lambda t\Gamma(k^m) \sim 1$. 
}
\label{fig:densityPlot}
\end{figure}

\begin{figure}
\includegraphics[width=0.99\linewidth]{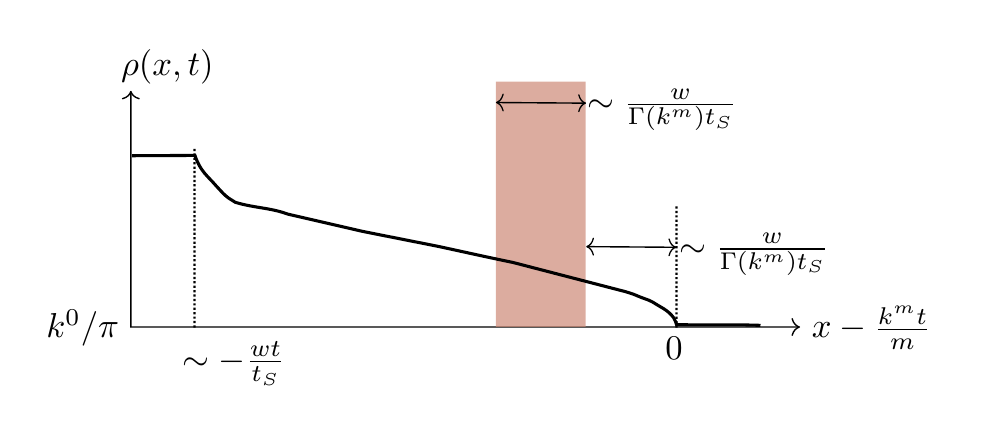}
\caption{
	Schematic of the density $\rho(x,t)$ as given by
	Eq.~(\ref{eq:combinedResult}). The ballistic form of $\delta \rho_{\rm
	high}$ in Eq.~(\ref{eq:deltarhohigh}) is preserved on a scale $\sim
	w/\Gamma(k^m)t_S$ from the tip of the shock. 
	There is a window of extent $\sim w/\Gamma(k^m) t_S$ where the contribution
	to $\delta \rho$ from $\delta \rho_{\rm low}$ given by
	Eq.~(\ref{eq:Fcorrection}) is significant.
	The overall size of the shock structure grows as $w t/t_S$.
}
\label{fig:smear}
\end{figure}

Having dispensed with ripples, simple geometric considerations dictate that the
integrand in Eq.~(\ref{eq:ffull}) is only nonzero for
\be
x-k(t-t') \leq x_+(t'),
\ee
which determines an inequality for $t'$ in Eq.~(\ref{eq:ffull})
\be
t'\geq t_1 \equiv \frac{x-kt}{k^m - k},
\ee
and so we can rewrite it as
\begin{multline}
f_{\rm low}(x,k,t)
=\\
\int_{t_1}^t \d t' e^{-(t-t')\Gamma(k)}
J\left(  (k^m-k)t_1 + kt',k,t' \right)
.
\label{eq:flowJ}
\end{multline}
Hereinafter, $x$ should be understood as a function of $\lambda$ and $t$.
Using these new variables, we may use the explicit expression for $\mathcal{W}$
given by Eq.~(\ref{eq:decayrate})
and discard subleading corrections in $\lambda$
to find
\begin{multline}
f_{\rm low}(x,k,t)
\approx
2 c
\Gamma(k^m) t_S
\left( \frac{k^m - k}{k^m - k^0} \right)^{11/2}
\left( \frac{k-k^0}{k^m-k^0} \right)^2 \\
\times
\int_{t_1}^t \frac{\d t'}{t_1}  \left(\frac{t_1}{t'} \right)^{6}
\sqrt{1-\frac{t_1}{t'}}
e^{-(t-t')\Gamma(k)}
e^{-t' \Gamma(k^m)}
.
\end{multline}
To determine the behaviour of this integral, it is crucial to know the
behaviour of the exponential inside the integrand. To make this clearer, we
introduce the dimensionless variable
\be
\tau \equiv \frac{t'-t_1}{t_1}.
\ee
In terms of this, we may write
\begin{multline}
f_{\rm low}
\approx
2 c
\Gamma(k^m) t_S
e^{-(t-t_1) \Gamma(k) - t_1 \Gamma(k^m)}
\left( \frac{k^m - k}{k^m - k^0} \right)^{11/2} \\
\left( \frac{k-k^0}{k^m-k^0} \right)^2
\int_0^{ \frac{t-t_1}{t_1}}  \frac{\d \tau \sqrt{ \tau}}{(1+\tau)^{13/2}} 
e^{-t_1 \tau \left( \Gamma(k^m) - \Gamma(k) \right)}
.
\label{eq:ftau}
\end{multline}
It will be helpful to also introduce a dimensionless variable $\gamma$ which
interpolates in the $k$-direction between the overhanging tip ($\gamma=0$) and the
background Fermi sea ($\gamma=1$), defined by
\be
k-k^0 = \left( k^m-k^0 \right)\left( 1-\lambda \right)\left( 1-\gamma \right).
\label{eq:gammadef}
\ee
Rewriting in terms of the variables $\lambda$, $\gamma$, $t$, the leading
behaviour of Eq.~(\ref{eq:flowJ}) may be approximated by
\begin{multline}
f_{\rm low}\left[ \lambda,\gamma,t \right]
\approx
2c
\Gamma(k^m) t_S \left[ \gamma+\lambda \right]^{11/2} \left( 1-\gamma \right)^2
e^{-t \Gamma(k^m)} \\
\times
\int_0^{\lambda/\gamma}
\d \tau
\frac{\sqrt{\tau}}{\left( 1+\tau \right)^{13/2}}
e^{-\tau \Gamma(k^m) t \gamma}
.
\label{eq:flowfinal}
\end{multline}
The contribution to the density coming from the ``high energy'' region i.e.
$k_-(x,t) < k < k_+(x,t)$ behaves as
\beA
\delta \rho_{\rm high}[\lambda,t]
&
=
\int_{k_-[\lambda,t]}^{k_+[\lambda,t]} \frac{\d k}{2\pi} W(x,k,t) \\
&\approx
(k^m - k^0) e^{-t \Gamma(k^m)}
\frac{t_S\sqrt{\lambda}}{t\pi}.
\label{eq:deltarhohigh}
\eeA
The contribution to the density coming from the ``low energy'' region is given
by integrating Eq.~(\ref{eq:flowfinal}) over $k^0<k<k_-(x,t)$, which
translates to
\be
\delta \rho_{\rm low} =
\frac{\left( k^m-k^0 \right)}{2\pi} \int_0^1 \d \gamma f_{\rm low}[\lambda,\gamma,t]
.
\label{eq:drholowdef}
\ee
Using Eq.~(\ref{eq:flowfinal}) and noticing that the dominant contribution to
$\delta \rho_{\rm low}$ comes from $\gamma \sim \mathcal{O}(1)$,
Eq.~(\ref{eq:drholowdef}) may be evaluated to leading order in $\lambda$, see
Appendix~\ref{app:corrections}, yielding
\be
\delta \rho^{\rm low}[\lambda,t]
\approx 
F(\lambda t\Gamma(k^m)) 
\delta \rho_{\rm high}[\lambda,t]
\label{eq:Fcorrection}
.
\ee
Here $\delta \rho_{\rm high}$ is given by Eq.~(\ref{eq:deltarhohigh}) and we
have used the function
\be
F(z) = 
\frac{8}{5\sqrt{z}} 
\int_0^{z} \d y \sqrt{y}
e^{-y},
\label{eq:Fdef}
\ee
which is plotted in Fig.~\ref{fig:F}.
We note that the form of $F(z)$ appears to be largely insensitive to the
particular form of $\mathcal{W}$.
This correction is maximal at $\lambda t\Gamma(k^m) \sim
1$.
We have argued that the ballistic result of Eq.~(\ref{eq:deltarhohigh}) is
modified with a contribution from ``lower energies'', giving
\begin{multline}
\delta\rho[\lambda,t]
\approx
(k^m-k^0)
e^{-t \Gamma(k^m)} \\
\times
\frac{t_S\sqrt{\lambda}}{t\pi}
\left[ 1+ F(\lambda t\Gamma(k^m) ) \right],
	\label{eq:combinedResult}
\end{multline}
which remains a monotonic function of $\lambda$, as shown in Fig.~\ref{fig:densityPlot}.
In dimensionful variables the shock wave preserves its form away from the
front over the time-independent scale $x_+(t) -x \sim \frac{w}{\Gamma(k^m)
t_S}$, while the entire shock wave structure (see Fig.~\ref{fig:smear})
expands linearly in time as $x_+(t) - x_-(t) \sim w \frac{t}{t_S}$.

\label{sec:Jconsistent}
With our expression for $f_{\rm low}$ in Eq.~(\ref{eq:flowfinal}), it is
possible to comment on the consistency of the assumption of the region
$[k_-(x,t),k_+(x,t)]$ being the dominant contribution to $J(x,k,t)$, as asserted in
Eq.~(\ref{eq:Jfhigh}). 
By evaluating the contribution to $J$ for momenta below $k_-(x,t)$, as
outlined in Appendix~\ref{app:Jconsistency}, one finds that the relative correction is not
asymptotically small for all times, with behaviour similar to
that of Eq.~(\ref{eq:Fcorrection}).
However, the correction is in fact numerically
small, on the order of $2\%$ at the largest, and so the approximation of
Eq.~(\ref{eq:Jfhigh}) appears to be consistent.

\section{Conclusions} \label{sec:conc}
For free fermions, introducing a density disturbance of width $w$ containing
$\Delta N\gg 1$ particles leads to the formation of a shock structure on a timescale
which may be estimated from classical mechanics as $t_S \sim \frac{w^2}{\Delta
N}$.
Semi-classical corrections in the form of quantum ripples exist in a region of
the front of the shock for all times. As quantified in
Section~\ref{sec:freefermions}, the size of this region grows linearly with
time, but with a parametrically small coefficient.
In dimensionful variables, the quantum ripples occur on the scale $x_+(t) - x
\sim \frac{wt}{t_S \left( \Delta N \right)^{2/3}}$.
The fraction of the shock for which ripples are significant is given
by $\lambda_{\rm cr} \sim \left( \Delta N \right)^{-2/3}$, and so most of the
shock is well-described classically.

Utilising this picture and disregarding ripples, one may now consider turning
on interactions.
Generic interactions lead to the decay of excitations high above the Fermi sea
at the expense of creating a large number of low-energy excitations.
These high-energy excitations have characteristic decay rate $\Gamma(k^m)$.
To establish continuity between the free and interacting pictures, we make the
restriction that $t_S \Gamma(k^m) \ll 1$:
the profile of the shock is manifestly unchanged for short times, as
interactions have barely ``turned on''.
Decay processes are only pertinent at $t \Gamma(k^m) \gtrsim 1$, well after the
shock has been established.

Our analysis of the linearised Boltzmann equation of Eq.~(\ref{eq:boltzmann})
yields the behaviour of the fermionic density at times $t\gg t_S$.
For any time $t \gtrsim \Gamma(k^m)^{-1}$ the deviations from a simple
exponentially decaying ballistic theory are given by the function $F(\lambda t
\Gamma(k^m))$ in Eq.~(\ref{eq:combinedResult}).
This term is significant only for $\lambda \sim 1/\left[ t \Gamma(k^m)
\right]$.
In dimensionful variables, the kinetic corrections to the density profile
become important only at distances $x_+(t) - x \gtrsim w/\left[t_S \Gamma(k^m)
\right]$ away from the tip of the shock wave, illustrated in
Fig.~(\ref{fig:smear}).  We find that the corrections to the shape are small
at $x_+(t) - x \ll w/\Gamma(k^m) t_S$, albeit the density of fermions residing
in the shock wave is suppressed by the factor $e^{-t \Gamma(k^m)}$.

As the region of quantum ripples grows linearly with time, it eventually
overtakes the scale $w/\Gamma(k^m) t_S$.
It is meaningless to make claims on the modified shape of the classical profile
once quantum ripples have encroached upon this region.
The length-scale for ripples and the length-scale for kinematic corrections are
comparable for $t \Gamma(k^m) \sim \left( \Delta N \right)^{2/3}$.
This time-scale therefore determines when Eq.~(\ref{eq:combinedResult}) is no
longer legitimate, as quantum corrections are important.
Until this time, however, the main result of Eq.~(\ref{eq:combinedResult})
holds: the shape of the shock is modified and subject to exponential decay of
the magnitude with rate $\Gamma(k^m)$.
We note that by the time $t\sim(\Delta N)^{2/3}/\Gamma(k^m)$ at which the
``quantum'' distortion meets with the ``kinetic'' one, destroying the shape as
described by Eq.~(\ref{eq:combinedResult}), its amplitude becomes exponentially
small in the large parameter $(\Delta N)^{2/3}$ of the semi-classical theory.

A priori it is not clear how interactions should modify a shock wave.  We have
provided a picture motivated by a Boltzmann equation, where
we explicitly determine the shape of the propagating shock at times well after
the formation of the shock, including generic interactions.
Although $\Gamma(k^m)$ is given by a perturbative evaluation of Fermi's Golden
Rule for generic density-density interactions between spinless
fermions\cite{PKKG}, the strong momentum-dependence of $\mathcal{W}$ in
Eq.~(\ref{eq:decayrate}) is a consequence of the limited phase space for
scattering, which remains true even in the case of spinful fermions.
We conjecture that our observations should be quite generic, as the form of
$F(z)$ in Eq.~(\ref{eq:Fdef}) is not sensitive to the precise details of the
rate $\mathcal{W}$.
The main result of Eq.~(\ref{eq:Fcorrection}) exhibits the corrections to the
na\"{\i}ve picture of a kinematic shock with exponential decay.

It remains of interest to investigate if there is a direct connection with the
non-linear Luttinger liquid\cite{SIG} picture, in order to investigate this question in a
more general, non-perturbative context\cite{idrisovschmidt}.

\section{Acknowledgements}
We acknowledge support from the Yale Postdoctoral Prize Fellowship (TV) and NSF
DMR Grant No. 1603243 (LG).

\onecolumngrid
\appendix

\section{Evaluating the bulk profile}
\label{app:bulk}
The approach we take to understand the time-evolved behaviour of
Eq.~(\ref{eq:Wt0}) for times $t \gg t_S$ is to consider a quadratic profile
\be
k_F(x) = 
\begin{cases} 
k^m - \left( k^m - k^0 \right)\left( \frac{x}{w} \right)^2, & |x|< w \\ 
k^0, & |x| \geq w
\end{cases}.
\label{eq:quadprofile}
\ee
When focussing on the front of the shock only the curvature at the maximum of
the initial density perturbation should be important.
The time evolution of the Wigner function is given by $x \to x - \frac{\hbar k
t}{m}$, and at fixed $x,t$ the roots of $k_F\left( x-\frac{kt}{m} \right)-k$
determine the upper and lower boundaries of support of the overhanging section
of classical Wigner function.  The case of Eq.~(\ref{eq:quadprofile}) causes
these to be the roots of a quadratic polynomial
\be
k_F\left( x-\frac{kt}{m} \right) - k
=
-\left( k^m-k^0 \right)
\tau^2
\left( \kappa - \kappa_+ \right) \left( \kappa-\kappa_- \right)
,
\label{eq:kfkroot}
\ee
where
\beA
\kappa &\equiv \frac{k^m - k}{k^m - k^0},\\
\kappa_\pm &= \lambda \pm \frac{\sqrt{\lambda}}{\tau} , \label{eq:kpm} \\
\tau &\equiv \frac{t}{t_s}.  
\eeA
Here we have assumed that $\lambda \gg \frac{1}{\tau}$. 
To evaluate the density we must plug the time-evolved Eq.~(\ref{eq:Wt0})
into Eq.~(\ref{eq:rhodef}).
We wish to evaluate the excess density, defined as
\be 
\delta \rho(x,t) = \langle \Psi| \rho(x,t)|\Psi\rangle - \langle 0 | \rho(x,t)|0\rangle.
\ee
For the quadratic profile this is equivalent to restricting the $k$-integration
of Eq.~(\ref{eq:rhodef}) to the region for which $|x - \frac{kt}{m}| \leq
w$. 
In the new variables of Eq.~(\ref{eq:kpm}) this corresponds to
\be
\lambda - \frac{1}{\tau} \leq \kappa \leq \lambda+ \frac{1}{\tau}.
\label{eq:kbounds}
\ee
This window is sufficiently large to capture smearing of the Wigner function in
the $k$-direction even for non-quadratic profiles.
Integrating $\kappa$ between the bounds of Eq.~(\ref{eq:kbounds}) yields 
\be
 \delta \rho[\lambda,t]
 ={\left( k^m - k^0 \right)}
\int_{\lambda-\frac{1}{\tau}}^{\lambda + \frac{1}{\tau}}
\frac{\d\kappa}{2\pi}
\int \frac{\d y}{2\pi i\left( y+i0^+ \right)} 
e^{i y\left( k^m-k^0 \right)\tau^2 \left( \frac{\kappa}{\tau^2}-\left( \kappa-\lambda \right)^2 \right)
- \frac{iy^3}{12} \frac{k^m-k^0}{w^2}}
.
\ee
Rewriting the quadratic polynomial explicitly as in Eq.~(\ref{eq:kfkroot}), and
introducing $y' \equiv y \left( \frac{k^m - k^0}{4w^2} \right)^{1/3}$ and
$\kappa' \equiv \kappa-\lambda$, gives
\be
{ \delta \rho[\lambda,t]}
={\left( k^m - k^0 \right)}
\int_{-\frac{1}{\tau}}^{\frac{1}{\tau}}
\frac{\d \kappa'}{2\pi}
\int \frac{\d y'}{2\pi i \left( y' + i0^+ \right)}
e^{-i y' (\Delta N)^{2/3} \tau^2 \left( \kappa' - \kappa_+ + \lambda \right)\left( \kappa' - \kappa_- + \lambda \right) - i y'^3 / 3}
,
\label{eq:intermediaryRho}
\ee
where we have identified $\Delta N = 2 w(k^m - k^0)$. 
We may now examine the $\tau\to\infty$ behaviour of
Eq.~(\ref{eq:intermediaryRho}).
Up to a prefactor, this is in fact a function of one variable:
$\lambda/\lambda_{\rm cr}$, where 
\be
\lambda_{\rm cr} = \frac{1}{\left( \Delta N
\right)^{2/3}}.
\ee
This may be seen by defining $p \equiv \kappa' \tau (\Delta N)^{1/3}$, using
the expressions for $\kappa_\pm$ from Eq.~(\ref{eq:kpm}), and extending the
range of the $p$-integration to infinity to find
\be
\delta \rho[\lambda,t]={\left( k^m - k^0 \right)}
\int_{-\infty}^{\infty}
\frac{\d p }{2\pi\tau (\Delta N)^{1/3}}
\int \frac{\d y'}{2\pi i \left( y' + i0^+\right)}
e^{-i y'  \left( p^2- {\lambda}/{\lambda_{\rm cr}} \right) - i y'^3 / 3}
.
\label{eq:bulkappinit}
\ee
Taking spatial derivatives of Eq.~(\ref{eq:bulkappinit}), and making the
substitutions (assuming $\lambda>0$)
\be
y=\frac{1}{2} \left( u-v \right) \sqrt{\frac{\lambda}{\lambda_{\rm cr}}}
,
\qquad
p=\frac{1}{2} \left( u+v \right) \sqrt{\frac{\lambda}{\lambda_{\rm cr}}},
\ee
gives the form
\be
\partial_x \rho[\lambda,t]
=
\left( k^m - k^0 \right)
\frac{\lambda}{2\lambda_{\rm cr}} \frac{(\Delta N)^{1/3}}{ w \tau^2} \left\vert
\int \frac{\d u}{2\pi}e^{\frac{i}{2} \left( \lambda/\lambda_{\rm cr} \right)^{3/2} \left[  u - u^3/3 \right]} \right\vert^2,
\ee
which is valid at any $\lambda/\lambda_{\rm cr}$.
Using the definitions of $\lambda$, $\lambda_{\rm cr}$, $\tau$, $\ell(t)$ (of
Eq.~(\ref{eq:elldef})), and $\Delta N$, this is equivalent to
Eq.~(\ref{eq:dxrhoairy}).
The leading asymptote for $\lambda/\lambda_{\rm cr} \gg 1$ is
\be
 \delta \rho[\lambda,t]
\approx
\left( k^m - k^0 \right)
\left[ 
\frac{\sqrt{\lambda}}{\pi \tau}
+
\frac{ \sin \left( \frac{2}{3}  (\Delta N) \lambda^{3/2} \right)}{2\pi\tau\lambda \Delta N}
 \right]
.
\ee

\section{Determining corrections in ``low energy'' region}
\label{app:corrections}

We begin from the expression for $f_{\rm low}$ of Eq.~(\ref{eq:flowfinal}):
\be
f_{\rm low}[\lambda,\gamma,t]
\approx
2c
\Gamma(k^m) t_S e^{-t \Gamma(k^m)} 
\left[ \gamma+\lambda \right]^{11/2} 
\left( 1-\gamma \right)^2 
\int_0^{\lambda/\gamma} \frac{\d \tau \sqrt{\tau}}{\left( 1+\tau \right)^{13/2}}
e^{-\tau t \Gamma(k^m) \gamma}
.
\ee
The contribution to the density, given by integrating over $\gamma$ as in
Eq.~(\ref{eq:drholowdef}), relative to $\delta \rho_{\rm high}$ is given by
\be
\frac{\delta \rho_{\rm low}[\lambda,t]}{\delta \rho_{\rm high}[\lambda,t] } 
=
\frac{
c
\Gamma(k^m) t }{\sqrt{\lambda}}
\int_0^1 \d \gamma 
\left[ \gamma+\lambda \right]^{11/2} 
\left( 1-\gamma \right)^2 
\int_0^{\lambda/\gamma} \frac{\d \tau \sqrt{\tau}}{\left( 1+\tau \right)^{13/2}}
e^{-\tau t \Gamma(k^m) \gamma}
.
\label{eq:lowhigh}
\ee
We will show that this integral is dominated by $\gamma \sim \mathcal{O}(1)$.
In this case we may approximate Eq.~(\ref{eq:lowhigh}) by setting $\lambda=0$
in the lower integration limit and in $\left[ \gamma+\lambda \right]^{11/2}$,
and dropping the $(1+\tau)^{13/2}$ denominator, which will be
$1+\mathcal{O}(\lambda)$.
This then yields the simpler expression
\be
\frac{\delta \rho_{\rm low}[\lambda,t]}{\delta \rho_{\rm high}[\lambda,t] } 
\approx
\frac{
c
\Gamma(k^m) t }{\sqrt{\lambda}}
\int_0^1 \d \gamma \,
 \gamma^{11/2} 
\left( 1-\gamma \right)^2 
\int_0^{\lambda/\gamma} \d \tau \sqrt{\tau}
e^{-\tau t \Gamma(k^m) \gamma}
.
\label{eq:lowhigh1}
\ee
Eq.~(\ref{eq:lowhigh1}) depends only on the parameter $\lambda t\Gamma(k^m)$, and may be simply rewritten as 
\beA
\frac{\delta \rho_{\rm low}[\lambda,t]}{\delta \rho_{\rm high}[\lambda,t] } 
&
\approx
F(\lambda t \Gamma(k^m) ), \\
F(z)
&
=
\frac{8}{5\sqrt{z}}
\int_0^{z} \d y \sqrt{y}
e^{-y}
.
\label{eq:lowhighapprox}
\eeA
We will now justify this procedure.
First, we split the $\gamma$ integral into two regions, focussing first on the
region $\gamma<\lambda$:
\be
I_1 \equiv
\frac{
c
\Gamma(k^m) t }{\sqrt{\lambda}}
\int_0^\lambda \d \gamma 
\left[ \gamma+\lambda \right]^{11/2} 
\left( 1-\gamma \right)^2 
\int_0^{\lambda/\gamma} \frac{\d \tau \sqrt{\tau}}{\left( 1+\tau \right)^{13/2}}
e^{-\tau t \Gamma(k^m) \gamma}
.
\ee
By changing integration variables it is clear that to leading order in
$\lambda$ this is given by
\beA
I_1 & = \lambda^5 G(\lambda t\Gamma(k^m)), \\
G(k) &\equiv c k\int_0^1 \d x \int_0^1 \d y \frac{\sqrt{ y} (1+x)^{11/2} e^{- k x}}{\left( 1+y/x \right)^{13/2} x^{3/2}} ,
\eeA
where $G(k) \sim k$ for $k\ll1$, and is bounded by a constant for all $k>0$.
Turning now to the region $\gamma >\lambda$:
\be
I_2 \equiv 
\frac{c
\Gamma(k^m) t }{\sqrt{\lambda}}
\int_\lambda^1 \d \gamma 
\left[ \gamma+\lambda \right]^{11/2} 
\left( 1-\gamma \right)^2 
\int_0^{\lambda/\gamma} \frac{\d \tau \sqrt{\tau}}{\left( 1+\tau \right)^{13/2}}
e^{-\tau t \Gamma(k^m) \gamma}
.
\label{eq:I2form}
\ee
It is helpful to introduce the change of variables $z\equiv \tau t \Gamma(k^m)
\gamma$, and explicitly include $c$ from Eq.~(\ref{eq:decayrate}), such that
\be
I_2=
\frac{168}{\sqrt{\lambda t \Gamma(k^m)}}
\int_\lambda^1 \d \gamma 
\frac{\left[ \gamma+\lambda \right]^{11/2}}{\gamma^{3/2}}
\left( 1-\gamma \right)^2 
\int_0^{\lambda t \Gamma(k^m)} \frac{\d z \sqrt{z}}{\left( 1+\frac{z}{\Gamma(k^m) t \gamma} \right)^{13/2}}
e^{-z}
.
\ee
The leading behaviour in $\lambda$ is given by setting $\lambda=0$ in both the fractional power term and the lower limit of the $\gamma$ integral, as well as discarding the denominator in the $\tau$ integral, giving
\be
I_2
=
F(\lambda t \Gamma(k^m)) +\mathcal{O}(\lambda), 
\label{eq:I2final}
\ee
with $F(z)$ given by Eq.~(\ref{eq:lowhighapprox}).
The scale for $I_1$ to be comparable to $I_2$ is $\Gamma t\lambda
\sim \lambda^{-10}$. 
For decay to be significant we require $\Gamma(k^m) t \gtrsim1$, and as
$\lambda \ll 1$ we are justified in considering only the contribution from $I_2$.
Putting all this together allows us to write the leading contribution as
\be
\delta \rho_{\rm low}[\lambda,t]
=
F(\lambda t \Gamma(k^m))
\delta \rho_{\rm high}  [\lambda,t]
,
\ee
with $\delta \rho_{\rm high}[\lambda,t]$ given by Eq.~(\ref{eq:deltarhohigh}).
We note that the form of $\mathcal{W}$ determines the particular numerical
coefficients and powers appearing in the above expressions, but the form of the
integral of $F$ is insensitive to this.
Indeed, the factors of $1/(1+\tau)$ we neglect in the integrand of
Eq.~(\ref{eq:I2form}) come from the $(k-p)$ dependence of $\mathcal{W}$ in
Eq.~(\ref{eq:decayrate}). This dependence arises from a combination of matrix
elements and density-of-states of low-energy excitations. The density-of-states
is small even in the presence of spin, and so it is possible that the form of
$F$ survives even in the case of weakly-interacting spin-$\frac{1}{2}$
fermions.

\section{Consistency of approximation}
\label{app:Jconsistency}

We wish to understand if the assumption of Eq.~(\ref{eq:Jfhigh}) is consistent.
In order for this to be the case the contribution to the source term from the
``low-energy'' region should be small compared to that of the ``high-energy''
region.
This entails examining
\be
\frac{J_{\rm low}(x,k,t)}{J_{\rm high}(x,k,t)}
=
\frac{
	\int_k^{k_-(x,t)} \d p f(x,p,t) \mathcal{W}_{p\to k} 
}
{
	\int_{k_-(x,t)}^{k_+(x,t)} \d p f(x,p,t) \mathcal{W}_{p\to k} 
}
.
\ee
We can rewrite this by using the expression for $\mathcal{W}$ from
Eq.~(\ref{eq:decayrate}), the dimensionless variable $\gamma$ introduced in
Eq.~(\ref{eq:gammadef}), and the same approximation for the denominator as in
Eq.~(\ref{eq:avg}) to give
\be
\frac{J_{\rm low}[\lambda,\gamma,t]}{J_{\rm high}[\lambda,\gamma,t]}
=
\frac{
	\int_0^\gamma \d \gamma'  f[\lambda,\gamma',t] \left( 1-\frac{\gamma'}{\gamma} \right)^5
}
{
	2 \sqrt{\lambda} t/t_S e^{-t \Gamma(k^m)}
}
.
\ee
By applying the same approximation technique as in Appendix~\ref{app:corrections}, one
may obtain that the leading (in $\lambda$) relative correction to the source
term is given by
\be
\frac{J_{\rm low}[\lambda,\gamma,t]}{J_{\rm high}[\lambda,\gamma,t]}
=
\frac{\gamma^5 }{165} \left( 22 -20 \gamma + 5\gamma^2 \right)
F(\lambda t \Gamma(k^m))
+\mathcal{O}(\lambda),
\ee
where $F(z)$ is the same as in Eq.~(\ref{eq:lowhighapprox}).
We observe that although the corrections are $\mathcal{O}(1)$ for $\lambda t
\Gamma(k^m) \sim 1$, they are nonetheless numerically small, with the
largest corrections being below $2\%$.

\end{document}